\newcommand{\tsecompldate}{9th April 1999}

\documentclass[12pt,a4paper]{article}

\usepackage{amsfonts}

\newcommand{\vol}[1]{{\bf #1}}

\newcommand{\ttitle}[1]{{\it #1}}
\newcommand{\tpretitle}[1]{}
\newcommand{\prenumber}[1]{Report No. #1}

\newcommand{\tref}[1]{(\ref{#1})}

\newcommand{\tnotpre}[1]{#1}

\newcommand{\tpre}[1]{}
\newcommand{\tprenote}[1]{}

\newcommand{\tnote}[1]{}
\newcommand{\tcomment}[1]{}

\newcommand{\href}[2]{#2}
\newcommand{\eprint}[1]{{\tt #1}}

\newcommand{\tsedevelop}[1]{{}}

\def\tsetrue{T} \def\tsefalse{F} 

\let\tsepaper=\tsefalse   
\let\tsenoteon=\tsefalse   
\let\tseletter=\tsetrue  
\let\tsedevon=\tsefalse    
\if\tsetrue\tsepaper  \else
 \fi

\if\tsetrue\tsedevon 
\usepackage{showkeys} 
\renewcommand{\tsedevelop}[1]{{#1}}
\fi

\if\tsefalse\tsepaper

\fi

\if\tsetrue\tseletter 
\fi

\renewcommand{\vol}[1]{{\bf #1}}

\renewcommand{\ttitle}[1]{{\it #1}}
\renewcommand{\tpretitle}[1]{}
\renewcommand{\prenumber}[1]{Report No. #1}

\if\tsefalse\tsepaper 
\renewcommand{\tnotpre}[1]{}
\renewcommand{\tpre}[1]{#1}
\renewcommand{\href}[2]{{#2}{}}
\renewcommand{\eprint}[1]{\href{http://xxx.soton.ac.uk/abs/#1}{{\tt #1}}}
\renewcommand{\tpretitle}[1]{{\em #1},}
\fi

\if\tsetrue\tsenoteon 
\renewcommand{\tnote}[1]{\footnote{#1}}
\renewcommand{\tcomment}[1]{#1}
\fi

\newcommand{\half}{\frac{1}{2}}
\newcommand{\bea}{\begin{eqnarray}}
\newcommand{\eea}{\end{eqnarray}}
\newcommand{\beq}{\begin{equation}}
\newcommand{\eeq}{\end{equation}}
\newcommand{\beqn}{\[}
\newcommand{\eeqn}{\]}

\newcommand{\dK}{ \int_0^\infty dK }
\newcommand{\dDk}{ \int \frac{d^Dk}{(2\pi)^{D}} \;}
\newcommand{\ddk}[1]{ \int \frac{d^{#1}k}{(2\pi)^{#1}} \;}

\newcommand{\DET}{{\rm DET}}

\newcommand{\calV}{{\cal V}}

\begin{document}

\renewcommand{\thefootnote}{\fnsymbol{footnote}}

\tpre{\begin{flushright} {\tt Imperial/TP/97-98/33} \\
\eprint{hep-th/9803184}
\\
To appear Phys.Lett.B.
\\
23rd March 1998, final corrections 8th
April 1999
\\
Present version \tsecompldate \\ \tsedevelop{ (LaTeX-ed on \today
) \\}
\end{flushright}
\vspace*{1cm} }

\begin{center}
{\Large\bf Regularization schemes and the multiplicative
anomaly}\\
\tpre{\vspace*{1cm} } {\large T.S.
Evans\footnote{email: \href{mailto:T.Evans@ic.ac.uk}{{\tt
T.Evans@ic.ac.uk}}\tnotpre{, \tsecompldate}\tpre{, WWW:
\href{http://euclid.tp.ph.ic.ac.uk/links/time}{\tt
http://euclid.tp.ph.ic.ac.uk/\symbol{126}time} }}}
\\
\tpre{\vspace*{1cm}}
\href{http://euclid.tp.ph.ic.ac.uk/}{Theoretical Physics},
Blackett Laboratory, Imperial College,\\
Prince Consort Road, London, SW7 2BZ,  U.K.
\tnotpre{\\ Tel: U.K.-71-594-7837,
Fax: U.K.-71-594-7844 (or -7777) \\
PACS: 11.10.Gh,  02.30.Cj,  02.30.Tb
\\
Key Words: zeta-function, Wodzicki residue, regularization}
\end{center}

\begin{abstract}

Elizalde, Vanzo, and Zerbini have shown that the effective action
of two free Euclidean scalar fields in flat space contains a
`multiplicative anomaly' when $\zeta$-function regularization is
used. This is related to the Wodzicki residue. I show that there
is no anomaly when using a wide range of other regularization
schemes and that the anomaly can be removed by an unusual choice
of renormalization scales. I define new types of anomalies and
show that they have similar properties. Thus multiplicative
anomalies encode no novel physics. They merely illustrate some
dangerous aspects of $\zeta$-function and Schwinger proper time
regularization schemes.\tprenote{Contains additional material
not in the published version.}

\end{abstract}


{}\tnote{tnotes such as this not present in final version}

\renewcommand{\thefootnote}{\arabic{footnote}}
\setcounter{footnote}{0}


In QFT (quantum field theory) one often encounters terms of the
form $ \ln\det ( \Delta^{-1} )  $ where $\Delta$ is the propagator
for some field.  This arises for example in one loop
approximations to effective actions.\tprenote{Derivatives of
such expressions can be related to the expectation values of the
field which have a more direct physical meaning.}  However, in QFT
this is a determinant of an infinite dimensional matrix, since
there are an infinite number of physical modes.  It is therefore
of great interest to all practitioners of QFT that the
``multiplicative anomaly'' $a(A,B)$ \cite{Ka,EVZ},\tprenote{see
also \cite{KV}} where \beq a(A,B) :=  \ln\det ( AB ) - \ln\det (A)
- \ln\det ( B ), \label{adef} \eeq need {\em not} be zero if $A,B$
are infinite matrices.  The term anomaly refers to the failure of
a standard algebraic expression rather than its traditional use in
QFT in relation to symmetries.  As usual in QFT the individual
terms in \tref{adef} are naively infinite, so one must regulate
the ultra-violet before any serious discussion. I will indicate
that some regularization scheme, labelled $R$, has been used by
adding a subscript $R$, e.g.\ $a_R, [\ln\det(A)]_R$.

So far the anomaly has only been considered using $\zeta$-function
regularization \cite{EVZ}.\tprenote{see \cite{EVZ} for
citations\tcomment{, just above equation (1.9)}. In some sense I
am extending the strict definition of the anomaly from one in
terms of $\zeta$-function regularization to one using a wider
class of regularization schemes.} In particular, if $A$ and $B$
are elliptic operators and $\zeta$-function regularization is used
then the anomaly $a_\zeta$ is not generally zero. This $a_\zeta$
is related to the Wodzicki residue \cite{Wo,Ka}\tprenote{In
\cite{Wo} see sec 7.13, pp.176. In \cite{Ka} see section 6.5,
pp.225-226 and compare his eqn.(9) against my (1).  Also note the
comments  in section 1.2, equn.(4) which cites Wodzicki's 1984
thesis for a residue formula.  See also \cite{KS} (1.4), and
Wodzicki \cite{Wo2}.} which has been used in the non-commutative
geometric approach to particle physics \cite{Co}.\tnote{See Connes
\cite{Co}, page 307.} Recently Elizalde, Vanzo and Zerbini
\cite{EVZ} showed that the multiplicative anomaly in
$\zeta$-function regularization produces surprising results in the
simplest of QFT problems.  For instance, for two real free
relativistic scalar fields of mass $m_1,m_2$ respectively, in flat
$D$-dimensional Euclidean space-time of space-time volume
$\calV_D$, $D$ an even positive integer, \cite{EVZ}\tprenote{See
equation (2.20) \cite{EVZ}.} gives the anomaly as \bea
a_\zeta(\Delta_1^{-1},\Delta_2^{-1}) &=& \calV_D
\frac{(-1)^{D/2}(m_1^2 - m_2^2)^{D/2}}{D \; C(D)} .
(\Psi(1)-\Psi(D/2))  , \label{azres} \eea where $\Delta_j^{-1}=
-\partial^2 + m_j^2$, and $C(d) =
(4\pi)^{d/2}.\Gamma(d/2)$.\tprenote{e.g. see Veltman \cite{Ve}
Appendix C p233.   The subscript on the $a$ denotes that
$\zeta$-function regularization was used.}

At first sight this result is surprising --- how can two {\em
free} scalar fields mix in this way?  In fact in QFT such free
fields {\em could} be mixed by choosing a vacuum with a condensate
mixing the two fields\tprenote{For example see \cite{Str}.} and
perhaps the calculation is revealing this possibility.  For
instance, neutrino mixing and kaon oscillations can be modelled
using a simple quadratic Lagrangian with no mixing/interaction
terms if one chooses a non-standard vacuum
\cite{BV,ABIV}.\tprenote{See also \cite{ABIV2,BHV}.}

In this letter, I ask why such anomalies have not been noted in
standard QFT texts. To do this I extend the definition of the
anomaly of \cite{EVZ} from one defined in terms of
$\zeta$-function regularisation to a wide variety of popular
regularization schemes. I also define two new types of anomaly to
illustrate that there are a large number of such multiplicative
anomalies. One conclusion is that the existence of the
multiplicative anomalies depends on the type of regularization
scheme used. \tpre{The analysis here suggests a general
principle, namely that only mass-dependent regularization schemes
can produce anomalies. Thus simple sharp momentum cutoffs and
dimensional regularization usually have no multiplicative
anomalies.  On the other hand, $\zeta$-function regularization and
all the other Schwinger proper time regularizations \cite{Ball}
change the propagators so that they are mass-dependent schemes
which leads to anomalies. Since physics must be independent of the
regularization scheme, (though this will only be approximately
true in a general approximation scheme \cite{PDG} \tpre{ and
\cite{St}}) and anomalies are zero in some schemes, this shows
that multiplicative anomalies contain no new physics. They are
merely a technical complication of some regularization schemes. At
the very least they are equivalent to a finite shift in counter
terms and I will indeed show that they are equivalent to a shift
in the renormalization scales. Thus this analysis suggests that
$\zeta$-function regularization\tprenote{and the related
Schwinger proper time regularizations} is a difficult scheme to
use in flat-space time where changes to the analytic structure of
the propagators are involved e.g.\ a self-energy shift. Yet
$\zeta$-function regularization has been of great use in curved
space-time, so the role of the multiplicative anomaly in such
calculations needs to be examined.}

Consider the example of two free scalar fields.  One formal
definition of the partition function, $Z$, is\tprenote{I use
standard path integral methods for illustration purposes but
canonical methods give the same results}
\bea
Z & = & \int D\phi_1
D\phi_2 \; \exp \{ -\int d^Dx \;
 ( \phi_1 \Delta_1^{-1} \phi_1 +  \phi_2 \Delta_2^{-1} \phi_2 ) \}
= e^{-W_1} . e^{-W_2} , \label{Zfactored} \\ W_j &=& \half \ln\det
( \Delta_j^{-1}/\mu_j^2 )  ,
\nonumber
\eea
where I have
anticipated renomalisation by including renormalization scales
$\mu_j$.  The partition function may also be written as in terms
of a vector of fields $(\vec{\Phi})_j = \phi_j$ with mass matrix
$M^2_{ij}=m^2_i\delta_{ij}$ ($i,j=1,2$),
\bea
Z& = & \int
D\vec{\Phi}  \; \exp \{ -\int d^Dx \; {\Phi}_i( -\partial^2
\delta_{ij}  + M^2_{ij} )\Phi_j \}
\label{Zmatrix}
\\
&=& [ \DET \{
(-\partial^2 \delta_{ij}  + M^2_{ij}) /\mu_{12}^2 \} ]^{-1/2}
 = \exp( - W_{12})
\label{W12full}
\tpre{\\ &=& [ \det \{ ( (\partial^2)^2 -
\partial^2 (m_1^2 + m_2^2)  + m_1^2 m_2^2 )/ \mu_{12}^4 \}
]^{-1/2}}
\\
W_{12} &=& \half   \ln\det \left(  \{ (\partial^2)^2 -  \partial^2
(m_1^2 + m_2^2) + m_1^2 m_2^2 \} / \mu_{12}^4 \right)
\label{W12st}
\eea
Note that the direct calculation of the
gaussian in a product space of space-time and
fields ($x$ and $i$ indices respectively) produces the result
\tref{W12full}, with the determinant over this product space denoted
by $\DET$. However, moving to a determinant just over
space-time indices, denoted by $\det$, is not usually trivial as
pointed out by Dowker \cite{Do}.   Here though, the matrix is
diagonal in the field indices and in this special case Dowker's
general formula confirms that \tref{W12st} follows from
\tref{W12full}.\tprenote{See appendix for further discussion of
this point.}

Finally one can formally factorize the operator and then the determinant
in $W_{12}$ of \tref{W12st}
to reproduce \tref{Zfactored}.  The anomaly
$a(\Delta_1^{-1},\Delta_2^{-1}) = 2W_{12} -2W_1 - 2 W_2$
expresses the  failure of this
factorisation. However, first one
must regulate these UV infinite expressions.

The standard regularization scheme of modern particle physics is
dimensional regularization\tprenote{In effect, dimensional
regularization achieves its regularization by merely modifying the
measure of the loop-momenta integrations $\int d^Dk$ by a factor
of $(k^2/\mu^2)^{-\epsilon}$, see \cite{Le} for details.}
\cite{Le,PDG} which gives
\bea
2 [W_j]_\epsilon &=& {\calV_D}
\ddk{d_j} \ln \left( \frac{k^2 + m_j^2}{\mu_j^2}\right)
\label{Wjb}
\\
{} 2 [W_{12}]_\epsilon &=& {\calV_D} \ddk{d_{12}} \ln \left(
\frac{( k^2 + m_1^2)( k^2 + m_2^2)}{\mu_{12}^4} \right)
\label{W12b}
\eea
where $d_x=D-2\epsilon_x$, $D \in \mathbb{Z}^+$,
$\forall x \in \{ 1,2,12 \} $, and we use an $\epsilon$ subscript
to label quantities calculated in dimensional regularization. For
the simple problems considered here, dimensional regularization is
merely changing the loop integration to
\beq
\left[ \ddk{d}
\right]_\epsilon = \frac{1}{ C(d) } \dK
K^{D/2-1} \frac{K}{ \mu^{2} }{-\epsilon}
\label{drdef}
\eeq
where $K=k^2$.  For later use,
note that it is clearly a mass independent regularization scheme
provided the renormalization scale, $\mu$, remains an independent
parameter. We quickly see that the anomaly in dimensional
regularization is
\beq
a_\epsilon (\Delta_1^{-1},\Delta_2^{-1}) =
2[W_{12}]_\epsilon - 2[W_1]_\epsilon - 2[W_2]_\epsilon =0
\label{aeres}
\eeq
provided we choose
$\epsilon_{12}=\epsilon_{1}=\epsilon_{2}$ and $\mu_{12}=\mu
_{1}=\mu_{2}$.\tprenote{The equality of the $\epsilon$'s ensures
that $a_\epsilon$ is finite, agreeing with naive power counting
suggests it is finite.} This result follows because once
regularized we can manipulate the integrals freely and it is
legitimate  to combine the logarithms of the integrands in
\tref{Wjb} and \tref{W12b} into a single logarithm under one
integration.  In a similar way we can see that {\em any}
regularization scheme which works by altering the range or measure
of loop momenta integrations (denoted by LR) using a single
renormalization scale $\mu$ has no anomaly, $a_{{\rm LR}} =0$, as
all the integrals involved are well behaved. \tpre{\bea
\lefteqn{a_{{\rm LR}} (\Delta_1^{-1},\Delta_2^{-1}) :=
2[W_{12}]_{{\rm LR}} -2[W_1]_{{\rm LR}} -2[W_2]_{{\rm LR}} }
 \\
&=& {\calV_D} \left[ \dDk \right]_{{\rm LR}} \left\{ \ln \left(
\frac{( k^2 + m_1^2)( k^2 + m_2^2)}{\mu^4}\right) - \ln
\left(\frac{ k^2 + m_1^2}{\mu^2}\right) - \ln \left(\frac{ k^2 +
m_2^2}{\mu^2}\right) \right\} \nonumber \\ &=& 0 \eea } For the
case of a sharp momentum cutoff, there can be no doubt that the
above analysis is correct.  Since dimensional regularization has
some technical complications, \tpre{e.g.\ see
\cite{Na},\tprenote{See p.89 \cite{Na}.}} I will give a more
detailed analysis below.

Before that, consider a scheme where $\ln(A) \rightarrow (-sA^s)^{-1}$ \cite{DC}
which I will call $s$-regularization.  This is one of the family of
Schwinger proper time regularizations \cite{Ball}.  Thus one defines
\bea
2[W_j]_s &=& -{\calV_D} \frac{1 }{s_j} \dDk
\left(\frac{\mu_{j}^{2} }{( k^2 + m_j^2)}\right)^{s_j}   ,
\label{Wjs}
\\
{} 2[W_{12}]_s &=& -{\calV_D} \frac{2}{s_{12}} \dDk \left( \frac{
\mu_{12}^{4} }{( k^2 + m_1^2)( k^2 +m_2^2)} \right)^{s_{12}/2}  .
\label{W12s} \eea As a series about $s_x=0$, these expressions
have a simple pole, a constant term and higher order corrections
just like the small $\epsilon$ dimensional regularized
results.\tprenote{Note that this $s$- regularization scheme is
similar to, yet crucially different from, the `analytic
regularization' scheme \cite{Le} in that a single $s$ parameter is
used in \tref{W12s}.  Leibrandt  \cite{Le} says having one $s$
parameter per propagator is ``essential'' in `analytic
regularization', see after equation (1.3) of \cite{Le}, but
interestingly Delbourgo \cite{De} in section 5.2 does not mention
this.}  Note that the physical masses are clearly an integral part
of the regulatory mechanism making this a mass-dependent scheme.
The anomaly in the $s$-regularization scheme is then found to be
$a_s (\Delta_1^{-1},\Delta_2^{-1}) = 2[W_{12}]_s - 2[W_1]_s -
2[W_2]_s \neq 0 $. Again the integrals are all finite for non-zero
but small $s$ so the integrands can be combined.  However, the
result is obviously not zero because $2(AB)^{-s/2} \neq (A)^{-s}
+(B)^{-s}$. The whole family of Schwinger proper-time
regularizations \cite{Ball}\tnote{see Ball \cite{Ball} especially
appendix  A.2. Note that some members of the Schwinger proper-time
regularization family have the same name as other well known
regularizations, e.g.\ one is called dimensional regularisation,
but they are {\em not} the same as Ball carefully notes.} of which
$s$-regularization is the simplest example, give the same result.

To make contact with the $\zeta$-function
regularization scheme \cite{DC,Ball,EORBZ} used in \cite{EVZ} one
applies
\beqn
\zeta(s|A) = \frac{1}{\Gamma(s)} \int_0^\infty dt \; t^{s-1}
{\rm Tr} \{e^{-tA} \}
\eeqn
to \tref{Wjs} and \tref{W12s} to prove that
\bea
2[W_A]_s &=&
-\frac{1}{s}
\zeta(s | A )
=
-\frac{1}{s} \zeta(0 | A ) + 2 [ W_A]_\zeta + O(s), \; \; \; {}2[
W_A]_\zeta  := \left. \frac{d}{ds}  \zeta(s | A) \right|_{{s}=0}
\nonumber \eea where $\exp \{ -W_A \} = \int D\phi \exp \{ - \phi
A \phi \}$. Thus the $\zeta$-function method in this context is
equivalent to throwing away the diverging term of the
$s$-regularized expressions.\tprenote{In this sense the finite
nature of $Z$ and $W$ in the $\zeta$-function method is not very
special.  One can mimic this in dimensional regularisation  and
define a ``finite dimensional-regularization scheme'' by taking
the derivative with respect to $\epsilon$ of $\epsilon$ times the
usual dimensional regularized result.} \tpre{So $\zeta$-function
`regularisation' is to Schwinger regularisation schemes what the
MS (minimal subtraction) {\em renormalization} scheme is to
dimensional regularisation.} Since the divergent parts of the
$s$-regularization cancel in the anomaly $a_s$, it follows that
$a_s (\Delta_1^{-1},\Delta_2^{-1}) = a_\zeta
(\Delta_1^{-1},\Delta_2^{-1}) \neq 0$, in agreement with
\tref{azres} of \cite{EVZ}.

To allay any suspicions that there is some slight of hand hiding in
the manipulation of infinite quantities, I will present a
generalisation of a calculation in \cite{EVZ}.  Since dimensional
regularization alters the momentum integration, while
$s$- and $\zeta$- regularizations involve changes to the
propagators, I can combine these two techniques.  Thus I consider
the anomaly using a mixed ``$s\epsilon$-regularization'',
$a_{s \epsilon} (\Delta_1^{-1},\Delta_2^{-1}) =
2[W_{12}]_{s \epsilon} - 2[W_1]_{s \epsilon} - 2[W_2]_{s \epsilon}$,
where
\bea
{} 2 [W_j]_{s \epsilon} &=& -
{\calV_D} \frac{1 }{s_j} \mu_{j}^{2s_{j}+2\epsilon_j}
\ddk{d_j} ( k^2 + m_j^2)^{-s_j}   ,
\nonumber 
\\
{} 2 [W_{12}]_{s \epsilon}
 &=& -{\calV_D} \frac{2}{s_{12}}
\mu_{12}^{2s_{12}+2\epsilon_{12}} \ddk{d_{12}} \left( ( k^2 +
m_1^2)( k^2 +m_2^2) \right)^{-s_{12}/2} . \label{W12se} \eea To
calculate the anomaly I use the same approach as \cite{EVZ} and
assume that it is a function only of the mass difference, so I
must set $\mu_1=\mu_2$. Thus by taking the $m_2^2=0$ limit and
using standard integrals, \tpre{ \cite{GR}, \bea \int_0^\infty
dx \; x^{\mu-1} (1+x)^{-\nu} &=& \frac{ \Gamma(\mu)
\Gamma(\nu-\mu)}{ \Gamma(\nu)} , \label{Iint} \eea} I find the
following result for $D=4$ dimensions \bea a_{s \epsilon}
(\Delta_1^{-1},\Delta_2^{-1}) &=& \calV_4 \frac{(m_1^2-m_2^2)^2}{4
C(4)}
 \left( \frac{1}{1+r} (\Psi(2)-\Psi(1)) +
\ln\left(\frac{\mu_{12}}{\mu_1}\right) \right) ,
\label{ase}
\eea
where $r := \lim_{s,\epsilon \rightarrow 0} ( \epsilon/s)$.
I also choose $\epsilon=\epsilon_x$, $s=s_x$
$\forall x \in \{ 1,2,12 \}$ as this is
{\em essential} if the UV divergences are to
cancel.

Let us first set the renomalisation scales $\mu_x$ equal, as in
\cite{EVZ} and most other work.  The pure $s$- and $\zeta$-regulated
answers come from putting $\epsilon=0$ ($r=0$).
Thus we are not surprised that setting $\epsilon=0, \mu_{12}=\mu_1$
in \tref{ase} confirms the result \tref{azres} of \cite{EVZ} (apart
from a sign) where pure $\zeta$-function methods were used.
I also get the same result for $a_\zeta$ when using the Wodzicki residue
approach \cite{Ka,Wo,Co}.  At the
other extreme, setting $s=0$ ($r=\infty$) gives a pure dimensional
regulated answer, and indeed with all $\mu_x$ equal,
$a_{s\epsilon}=0$ confirming \tref{aeres}.  Thus the anomaly
depends on the regulation used.

Another rigorous check of these results is to calculate the
difference between derivatives of the $W$ functions.  This defines a new type of multiplicative
anomaly which I will call the
``$b$-anomaly''.  In the mixed $s\epsilon$-regularization this is
\bea
\frac{1}{4}b_{s \epsilon} (\Delta_1^{-1},\Delta_2^{-1})
&=& [\frac{\partial W_{12}}{\partial {(\delta m^2)}}]_{s \epsilon}
-[\frac{\partial W_1}{\partial {(\delta m^2)}}]_{s \epsilon}
+[\frac{\partial W_2}{\partial {(\delta m^2)}}]_{s \epsilon}
\nonumber 
\\
{} 4[\frac{\partial W_j}{\partial {(\delta m^2)}}]_{s \epsilon}
&=& {\calV_D} \mu_j^{2(s_j+\epsilon_j)} \ddk{d_j} \frac{1}{(k^2 +
m_j^2)^{1+s_j}} \tpre{= \langle \phi_j^2 \rangle}
\nonumber 
\\
{} 4[\frac{\partial W_{12}}{\partial {(\delta m^2)}}]_{s \epsilon}  &=&
\calV_D \mu_{12}^{2(s_{12}+\epsilon_{12})}
\ddk{d_{12}} 
\frac{(m_2^2-m_1^2)}{(k^4 + (m_1^2 + m_2^2) k^2 + m_1^2
m_2^2)^{1+s_{12}/2}} \tpre{=  \langle (\phi_1^2 - \phi_2^2)
\rangle}
\nonumber 
\eea where ${\delta m^2} = (m_1^2 - m_2^2)$.\tnote{These
derivatives are proportional to the expectation values of the
fields squared.} In this case one can do the small $\epsilon$ and
$s$ calculation with arbitrary $m_j^2$.  \tpre{From \cite{GR} I
find that\tprenote{First is Gradshteyn and Ryzhik \cite{GR}
equation (3.252.11), second is (8.711.1). Checked with MAPLE
22-3-97 for D=3.9, s=0.1, (and similar values) using
hypergeometric representation [AS] (8.1.2) for associated Legendre
polynomial, against MAPLE's direct numerical calculation. Expand
the second integral in the small parameters and do the resulting
integrals.  One may also do the $\epsilon$ expansion of the $P$
using \cite{AS} equation (8.6.9).} \bea \lefteqn{\int_0^{\infty}
dx \; \frac{x^{-\nu-1} }{(1+2\beta x + x^2)^{1/2 + \mu} } }
\nonumber \\ & =& \frac{(\beta^2-1)^{\nu/2} \Gamma(1+\mu)
\Gamma(\nu + 2 \mu +1) \Gamma(-\nu)}{\pi^{1/2}
\Gamma(1/2+\mu)\Gamma(1+2\mu)} \int_{-1}^{+1} dt
\frac{(1-t^2)^{\mu-1/2}}{(t+\beta(\beta^2-1)^{-1/2})^{- \nu}}  .
\nonumber \eea Expansion of the $t$ integral in the relevant small
parameters gives further integrals which can be calculated from
\cite{GR}. Using these} I find that \bea \lefteqn{b_{s \epsilon}
(\Delta_1^{-1},\Delta_2^{-1}) =} \nonumber \\ && \calV_D
\frac{(m_1^2-m_2^2)}{C(4)} \left[ \frac{1}{1+r} (\Psi(2)-\Psi(1))
- \ln \left( \frac{\mu_{12}^2}{\mu_1\mu_2} \right)
+
\frac{(m_1^2+m_2^2)}{(m_1^2-m_2^2)} \ln \left( \frac{\mu_1}{\mu_2} \right)
\right]
\label{bseres}
\eea
where again
$\epsilon=\epsilon_x$, $s=s_x$, $\forall x \in \{1,2,12 \} $ is
needed to ensure cancellation of the UV divergences.  If I use
a single renomalisation scale for the parts coming from the factored
calculation \tref{Zfactored} i.e.\ $\mu_1=\mu_2$, then the result is
a pure function of the mass difference.  In this limit \tref{bseres}
is then identical to
$2 \partial a_{s\epsilon}/ \partial {(\delta m^2)}$ as expected.

So far I have only looked at the results where all the
scales $\mu_x$ were set equal to a single independent scale parameter.
However, I can exploit the fact that my results for the anomalies are for arbitrary
scales, $\mu_x$.  Suppose I choose
$\mu_{12} = \mu_1 \exp \{ - (\Psi(2)-\Psi(1))/ (1+r) \}$,
so that in expressions like \tref{W12se} we can write
\beq
\mu_{12}^{2\epsilon_{12}+2s_{12}} =
\mu_{1}^{2\epsilon_{12}}.
\left( \mu_{1} e^{-(\Psi(2)-\Psi(1))} \right)^{2s_{12}}.
\label{mushift}
\eeq
This unusual choice for the renormalization scales removes the
anomaly for all hybrid $s\epsilon$-regularizations
as \tref{ase} shows.  Note that the shift is made purely in the $s$ sector
indicating it is the $s$- or $\zeta$-regularization which is
causing the problem.   The fact that these anomalies can be
removed with a shift in renormalization scales, and hence a change
in counter terms in the renormalized theory, is
a clear signal that the anomaly does not contain any novel physics

One can throw further light the nature of these multiplicative anomalies by
doing the reverse.  Consider the results \tref{ase} and \tref{bseres}
for pure-dimensional regularisation
($r=\infty$).  With a single scale $\mu$, the anomalies are zero.
However, suppose I define a mass-dependent  dimensional regularisation
scheme where $\mu^2$ for a given loop integral is set equal to the
average of the mass squared parameters in the integrands.
It is clear from the $\mu$ dependence of \tref{ase} and \tref{bseres} that the
anomalies are now non-zero except when $m_1 = m_2$, just as is found before with
the pure $s$- and $\zeta$-function regularisation methods.  Thus it appears that
multiplicative anomalies can always be interpreted as merely
reflecting use of different effective renormalization scales.\tnote{We
see that by choosing unequal renomalisation scales we can
eliminate the anomaly for {\em any} hybrid
$s\epsilon$-regularization scheme including the $\zeta$-function method.
 In particular the anomaly is zero when $\mu_1=\mu_2$ and $\mu_{12}$
is given by \tref{mushift}.  Only pure dimensional regularization
($s=0, \epsilon \rightarrow 0$)  has no anomaly when a single
\tpre{renormalization} scale $\mu$ is used.}

\tpre{One might ask if the same quantities are being compared in
the different schemes. The fact that the
$s\epsilon$-regularization scheme smoothly interpolates between
$\zeta$-function and dimensional regularization results shows that
I have a consistent definition of $[W]_R$ and $a_R$. The
differences obtained in different regularization schemes are
merely the expected variation found when considering unphysical
objects. It is well known that such shifts can be absorbed by a
redefinition of counter terms and this can be seen in the ability
to remove the anomalies by changing the renormalization scales
$\mu$.}

These results can be summarized by a set of three necessary
criteria for multiplicative anomalies to be present.  Firstly,
the unregulated terms of the anomaly must contain infinities.
Secondly, a mass-dependent regularisation scheme must be used, and
thirdly, different terms in the anomaly must contain different
masses. Presumably, anomalies appear whatever physical parameter, not just
mass, is involved.\tnote{The $s$-, $\zeta$-function, and general Schwinger proper time
regularizations satisfy the second criteria because they regulate
by altering the analytic structure of loop integrals.}

If the analysis above is correct, it suggests that there
are many other such anomalies associated with
mass-dependent regularisation schemes.  As a final check
consider a mass shift $\delta m^2=m_1^2-m_2^2$
to a free scalar field
of mass $m_2^2$.   I therefore define a new ``$\alpha$-anomaly'' to be
\bea
[\alpha (A,B)]_R &=& 2 [W_A]_R
- 2 \sum_{n=0}^\infty
\left[ \frac{b^n}{n!} \left. \frac{\partial^n}{\partial b^n} W_{B+b}
\right|_{b=A-B}
\right]_R
\nonumber 
\eea where $\exp \{ -W_A \} = \int D\phi \exp \{ - \phi A \phi
\}$.\tnote{This tests the expansion of $\ln \det (B+b)$ in terms
of $b/B$.}  For the simple mass shift I set $A=\Delta_1^{-1}$ and
$B=\Delta_2^{-1}$ so that $b=\delta m^2$.  Standard integrals
\tpre{\tref{Iint} \cite{GR}} give for $D=4$ in the
$s\epsilon$-regularization scheme \bea \alpha_{s\epsilon}
(\Delta_1, \Delta_2) &=& -\calV_4 \frac{ (m_1^2-m_2^2)^2}{C(4)}
\left( \frac{1}{1+r} + \frac{m_1^4}{ (m_1^2-m_2^2)^2} \ln
\left(\frac{\mu_1^2}{\mu_2^2}\right)  \right) \nonumber \eea where
\tpre{$r=\lim_{s,\epsilon}\epsilon/s$ and } I use the same $s$
and $\epsilon$ parameters in all integrals to get the divergences
to cancel.  This shows exactly the same behaviour as the $a-$ and
$b$-anomalies.  In particular with all the renormalization scales
set equal, the anomaly is zero {\em only} for $r=\infty$, pure
dimensional regularization.

The $\mu$ shift discussion shows that the anomalies have no novel
physical content, {\em pace} \cite{EVZ}. The fact that many simple
schemes have no anomalies suggests a more powerful argument in
support of this conclusion. It is a fundamental axiom of QFT that
the physics is independent of regularization scheme (in contrast
to {\em approximate} results which may depend on the scheme
\cite{PDG} \tpre{and \cite{St}}). Since I have shown that
anomalies are zero in some regularizations it follows that there
can be no novel physics in non-zero anomalies in other schemes.
Multiplicative anomalies are merely an annoying technical
difficulty present in certain regularisation schemes.  The
difference in the results for anomalies would then have to be
absorbed by differences in unphysical renormalization scales when
ensuring physical results are invariant. This is done regularly
when comparing calculations against real data \cite{PDG}. Here I
have explicitly demonstrated how anomalies can be removed by a
suitable redefinition of renormalization scales, supporting this
picture.\tnote{The alternative would be to assume that physics
does depend on the renormalization scale.  Different values of the
anomaly would then represent different physics and only one
regularization scheme would be physical --- only one value of $r$
in the mixed $s\epsilon$-regularization would be picked out.
\tpre{The simple models studied so far have not demonstrated
that there is any novel physics in anomalies.} As I have noted, a
careful matching on unphysical renormalization scales will be a
necessary part of such calculations.}

An important conclusion of this work is that that many different
multiplicative anomalies plague calculations using
$\zeta$-function and Schwinger proper time regularizations. In
these schemes the physical masses are entangled with the
regularisation scale since they regularize using factors such as
$[(k^2+m^2)/\mu^2]^{-s}$ \tpre{rather than
$[k^2/\mu^2]^{-\epsilon}$ of dimensional regularisation}. This
then ensures that the regulated integrands fail to obey many
algebraic identities naively satisfied by their unregulated
counterparts -- $\ln (AB) = \ln (A) + \ln (B)$ (the $a$ anomaly),
$ (AB)^{-1} = (1/A - 1/B)(1/(B-A))$ ($b$ anomaly), $\ln (1-A)^{-1}
= -\sum_n A^n/n$ ($\alpha$ anomaly).\tnote{Pauli-Villars
regularisation will also contain anomalies since it also mixes the
regularisation scale with the physical masses.} The key point is
that while a shift in renormalization scales $\mu$ can absorb the
anomalies in such schemes, the required shift seems to depend on
the expectation value being considered. So while \tref{mushift}
removes the $a-$ and $b$-anomalies, nothing like that removes the
$\alpha$ anomaly. On the other hand, regularizations which alter
the integration measure independent of physical parameters, e.g.\
sharp momentum cutoffs, dimensional regularization, space-time
lattices, leave the integrands unchanged if a single
renormalization scale $\mu$ is used, and then these never show an
anomaly. Thus there seems no good reason why one should choose
$\zeta$-function regularization with its multiplicative anomalies
for ordinary QFT problems.

Still, all regularization schemes have some drawbacks, e.g.\
dimensional regularization and curved space-time, momentum cutoffs
and gauge symmetry. The choice of regularization scheme is a
matter of personal taste and mathematical convenience given the
problem at hand. In problems involving curved space-time,
$\zeta$-function methods have proved to be most useful. The
crucial message is not that the multiplicative anomaly represents
novel physics but rather that when $\zeta$-function methods are
chosen the multiplicative anomaly {\em must} be taken into account
in such calculations to get the correct physical results. In this
sense it is of vital importance in many physical problems and
requires further study.


I would like to thank A.Filippi, and R.Rivers for useful discussions,
and G.Vitiello and Salerno University for stimulating some of them.
This work is the result of a network supported by the
\href{http://www.esf.org/}{European Science Foundation},
with additional support from
the European Commission through their Socrates and
Human Capital and Mobility ({\tt CHRX-CT94-0423}) programmes.


\newpage


\renewcommand{\thesection}{\Alph{section}}
\setcounter{section}{0}

\makeatletter
\renewcommand{\theequation}{\thesection.\arabic{equation}}
\@addtoreset{equation}{section} \makeatother \typeout{---
Equations labeled as (section.equation) ---}


\section{Evaluating the determinants}

Dowker \cite{Do} has made some important remarks regarding the
manipulation of determinants relevant to multiplicative anomalies.
Let me restate his point, extending his notation a little. Suppose
the space in which we are working can be split into two parts so
that each index of a matrix/operator is given by a pair $\alpha
i$. For quantum field theory the $\alpha$ represents the
N-different fields and $i$ represents the space-time index running
over $n^d$ values if we imagine working on a d-dimensional
space-time lattice with n lattice points in each direction. Using
capital letters to denote matrices/operators in this large space
we have, taking $N=2$ for simplicity,
\begin{equation}
M = \left( \begin{array}{cc} a & b
\\
c & d
\end{array}
\right) \label{Mdef}
\end{equation}
where I will use small letters to denote matrices in
$n^d$-dimensional space of space-time.  The point made by Dowker
is that
\begin{equation}
DET(M) \neq det(a) . det(d) - det(b) . det(c)
\label{Dpoint}
\end{equation}
where $DET$ is defined as the determinant in the full $\alpha i $
$(N n^d)$-dimensional space, while $\det$ denotes a
determinant taken only in the $i,j$ space-time indices.  As Dowker
points out, with $N=2$ and $n^d=2$, the general expression for
these matrices is a sum of 24 distinct terms on the left-hand side, while
the right-hand side is a sum of only 8 distinct terms confirming the
inequality given.

The focus of my letter is the case of two free scalar fields and I
will first discuss this problem. Later, I will turn to the case of
chemical potential which is discussed in another Elizalde et al
paper \cite{EFVZ} and which is also mentioned by Dowker.  In the
case of N free scalar fields, the propagator for $N$ fields can be
represented by the matrix $\Delta_{\alpha i, \beta j}$. As the
fields are free, the propagator is block diagonal in the field
indices, that is for $N=2$ it is of the form:-
\begin{equation}
\Delta = \left( \begin{array}{cc} a & 0
\\
0 & d
\end{array}
\right)  .
\end{equation}
The $a= - \partial^2 + m_1^2$ and $d= - \partial^2 + m_2^2$ are
$n^d$-dimensional matrices carrying only the space-time indices
$i,j$.  Now it is true in this special case that Dowker's general formula reduces
to:-
\begin{equation}
DET(\Delta) = det(a) . det(d) . \label{diagdet}
\end{equation}
so that in this special case equality in fact holds in \tref{Dpoint}.
An explicit check with $N=2$, $n^d=2$ confirms this analysis (e.g.
both sides are a sum of four terms) but it also follows from the
following general analysis.  Let
\begin{equation}
A = \left( \begin{array}{cc} a & 0
\\
0 & 1
\end{array}
\right) , \; \; \; D = \left( \begin{array}{cc} 1 & 0
\\
0 & d
\end{array}
\right)
\end{equation}
Then
\begin{eqnarray}
DET(\Delta) &=& DET(A.D) = DET(A).DET(D) \nonumber \\ &=&
[det(a).det(1) ]. [det(1).det(d)] = det(a).det(d) .[det(1)]^2
\label{diagderiv}
\end{eqnarray}
and for finite dimensional matrices I have reproduced
\tref{diagdet}.  The key point is that while Dowker makes a valid
point for general matrices \tref{Dpoint}, it is simplifies to
\tref{diagdet} for block diagonal matrices. In this case the
simpler expression \tref{diagdet} is true. This is the situation
encountered with free fields and this is why I used \tref{diagdet}
in going from \tref{W12full} to \tref{W12st}.

Incidently, the derivation \tref{diagderiv} again illustrates why
renormalisation scales are the the key to the problem . In
\tref{diagderiv} I have carefully shown all the factors of
$det(1)$ which are present.  These factors are in some sense
$1^\infty$ in quantum field theory and therefore a potential
source for infinite constants.  Also what do I mean by $1$?
Presumably this is one only with reference to some standard scale.
My point is that it is precisely such scales which are difficult
to understand in the zeta-function regularisation scheme and which
lead to the anomaly.

In the the case of chemical potentials, studied in \cite{EFVZ} and
mentioned by Dowker \cite{Do}, the propagator can be written with
non-zero entries in the off-diagonal blocks, $b,c\neq 0$ in the
terminology of \tref{Mdef}. So it appears that Dowker's point
might be of relevance. However, one can easily make such a
propagator block diagonal by working in the charge eigenstate
field basis (see \cite{TSEdal}) and then the proof given here
again shows once again that Dowker's point \tref{Dpoint} throws no
light on the multiplicative anomaly.

\if\tsenoteon\tsetrue
\section{Further details of calculations.}

These are the full equations for $W_s$ etc.
\bea
2[W_j]_s &=&
-\frac{1}{s_{j}}
\zeta(s_j | \Delta_j^{-1} / \mu_{j}^{2} )
=
-\frac{1}{s_j} \zeta(0 | \Delta_j^{-1}/ \mu_{j}^{2} ) + 2 [
W_j]_\zeta + O(s_j)
\label{Wjsb}
\\
{}2[ W_j]_\zeta &=&
\left. \frac{d}{ds_j}  \zeta(s_j | \Delta_j^{-1} / \mu_{j}^{2}) \right|_{{s_j}=0}
\\
{}2[W_{12}]_s &=&
-\frac{2}{s_{12}}
\zeta(s_{12}/2 | \Delta_1^{-1}\Delta_2^{-1} / \mu_{12}^{4} )
=
\frac{-2 }{s_{12}} \zeta(0 | \Delta_1^{-1}\Delta_2^{-1}/
\mu_{12}^{4} ) + 2 [ W_{12}]_\zeta + O(s_{12}) \label{W12sb}
\\
{}2[ W_{12}]_\zeta &=&
\left. \frac{d}{ds_{12}}
\zeta(s_{12} | \Delta_1^{-1} \Delta_2^{-1}/ \mu_{12}^{4}) \right|_{{s_j}=0}
\eea

The following can be used to do several of the integrals
\cite{GR}\tprenote{(3.381.4) in Gradshteyn and Ryzhik \cite{GR}}
\bea
\int_0^{\infty} dx \; x^{z-1}e^{-ax} &=& \Gamma(z) a^{-z}
\label{int2} \eea

One result used throughout can be proved using standard `Schwinger
proper time' trick \ref{int2}, or direct from Gradshteyn and
Ryzhik \cite{GR}, see (3.251.1) and (3.241.2), and which is used
again and again is \ref{Iint}.


Now I will compare this result to that obtained with regularization schemes which
alter the measure rather than the behaviour of the propagators,
$(2\pi)^{-D}d^Dk$, in the $I$ and $J$ integrals to some
$[(2\pi)^{-D}d^Dk]_R$.  This procedure includes
dimensional regularization
where I work in $d=D-2\epsilon$ dimensions and take $\epsilon$ to
zero at the end, simple cutoffs where I use
$(2\pi)^{-D}d^Dk \theta(\Lambda^2-k^2)$, and more complicated cutoffs
$(2\pi)^{-D}d^Dk \exp (-k^2/\Lambda^2)$, etc.
This ensures the integrals $I$ and $J$
are finite and thus I am free to manipulate the integrands without
fear.  Thus
\bea
J_R(m_1^2,m_2^2)
&=& \int[\frac{d^Dk}{(2\pi)^D}]_R \frac{1}{k^4 + (m_1^2 + m_2^2) k^2 + m_1^2 m_2^2}
\nonumber
\\
\tcomment{&=& \int[\frac{d^Dk}{(2\pi)^D}]_R
\frac{1}{(k^2+m_1^2)(k^2 + m_2^2)}\nonumber \\} &=&
\int[\frac{d^Dk}{(2\pi)^D}]_R \left[ - \frac{1}{k^2+m_1^2} +
\frac{1}{k^2 + m_2^2} \right] \frac{1}{m_1^2 - m_2^2} \label{jreg}
\eea and thus I see that \beq b_R (\Delta_1^{-1},\Delta_2^{-1}) =
0 \label{brdef} \eeq

Equation \ref{dW12def} may be
evaluated from Gradshteyn and Ryzhik \cite{GR}
equation (3.252.11).\tnote{Checked with MAPLE 22-3-97 for D=3.9, s=0.1,
(and similar values) using
hypergeometric representation [AS] (8.1.2) for associated
Legendre polynomial, against MAPLE's direct numerical calculation.}
\bea
\int_0^{\infty} dx \; \frac{x^{-\nu-1} }{(1+2\beta x + x^2)^{1/2 -
\mu} } &=& \left(\frac{\beta^2-1}{4}\right)^{\mu/2}
\frac{\Gamma(1-\mu) \Gamma(\nu - 2 \mu +1)
\Gamma(-\nu)}{\Gamma(1-2\mu)}  P^{\mu}_{\nu-\mu} (\beta)
\label{int1}
\eea
where $P^\mu_{\nu-\mu}$ is an associated Legendre
polynomial. So one can then use Gradshteyn and Ryzhik \cite{GR} (8.731.5) to
expand the arguments of the Legendre polynomial.

However, one may do the integrals laboriously in a manner very
similar to that used for the $\zeta$-regularized case, using the same
integrals \ref{int1} and \ref{int2}.  By using the result
\cite{AS}\tnote{\cite{AS} equation (8.6.9).  Checked numerically
against hypergeometric representation \cite{AS} equation (8.1.2)
for $\nu = -3.95/2$ and $z=1.1$ and similar values.}
\bea
P^{-1/2}_{\nu} (z) &=& \left(\frac{2}{\pi}\right)^{\half}
\frac{(z^2-1)^{-1/4}}{2\nu -1 }
\left[ (z+(z^2-1)^\half]^{\nu+\half}
-(z+(z^2-1)^\half]^{-\nu-\half}
\right]
\eea
I find that for $D=4$
\bea
J_\epsilon(m_1^2,m_2^2) &=& \frac{c_4}{2} \left[
\frac{1}{\epsilon} + \left(
-\frac{2c'_4}{c_4} -
\frac{2}{m_1^2-m_2^2} (m_1^2 \ln (\frac{m_1}{\mu_p}) )
-m_2^2 \ln (\frac{m_2}{\mu_p})
\right)
\right]
\\
I_\epsilon(m^2) &=& \frac{c_4 m^2}{2} \left[
-\frac{1}{\epsilon} +
\left(
\frac{2c'_4}{c_4}
- 2m^2 \ln (\frac{m_1}{\mu})
\right) \right]
\eea
where $c_d = d (\pi)^{-d/2} 2^{-1-d}/\Gamma(d/2+1)$\tnote{Bailin and
Love (7.15)} and $c'_x = d
(c_x) /dx$.
Putting these results in \ref{bdef} confirms that, for $D=4$,
$b_\epsilon=0$.

In exactly the same way one can look at a $\beta$-anomaly based on
$4\partial W_A / \partial (\delta m^2)$ for $m_2$ constant. A
direct calculation gives \bea \beta_{s\epsilon}(\Delta^{-1},\delta
m^2) &=& - \calV_4 \frac{m_1^2-m_2^2}{C(4)} \left( \frac{1}{1+r} +
+\frac{m_1^2}{m_1^2-m_2^2}\ln \left( \frac{\mu_1^2}{\mu_2^2}
\right) \right). \eea This is equal to $2 {\partial
\alpha_{s\epsilon}}{\partial {(\delta m^2)}}$ if the $\mu$'s are
assumed constant and thus consistent with the result
\ref{alphares}. This $b$-anomaly encodes a failure of the relation
$(1-A)^{-1} = \sum_n A^n$.

\fi

\end{document}